\documentstyle[preprint,aps]{revtex}
\begin{document}
\input{epsf}
\draft
\newfont{\form}{cmss10}
\newcommand{\e}{\varepsilon}
\renewcommand{\b}{\beta}
\newcommand{\unity}{1\kern-.65mm \mbox{\form l}}
\newcommand{\D}{D \raise0.5mm\hbox{\kern-2.0mm /}}
\newcommand{\A}{A \raise0.5mm\hbox{\kern-1.8mm /}}
\def\pmb#1{\leavevmode\setbox0=3D\hbox{$#1$}\kern-.025em\copy0\kern-\wd0
\kern-.05em\copy0\kern-\wd0\kern-.025em\raise.0433em\box0}
\def\D{\hbox{\hbox{${D}$}}\kern-1.9mm{\hbox{${/}$}}}
\def\kbar{\hbox{$k$}\kern-0.2truecm\hbox{$/$}}
\def\nbar{\hbox{$n$}\kern-0.23truecm\hbox{$/$}}
\def\pbar{\hbox{$p$}\kern-0.18truecm\hbox{$/$}}
\def\nhbar{\hbox{$\hat n$}\kern-0.23truecm\hbox{$/$}}
\newcommand{\dif}{\hspace{-1mm}{\rm d}}
\newcommand{\dil}[1]{{\rm Li}_2\left(#1\right)}
\newcommand{\diff}{{\rm d}}
\title{Two-dimensional QCD, instanton contributions and the perturbative
Wu-Mandelstam-Leibbrandt prescription}
\author{A. Bassetto, L. Griguolo }
\address{Dipartimento di Fisica "G. Galilei",
INFN, Sezione di Padova,\\
Via Marzolo 8, 35131 Padua, Italy}
\maketitle
\begin{abstract}
The exact Wilson loop expression for the pure Yang-Mills $U(N)$ theory
on a sphere $S^2$ of radius $R$
exhibits, in the decompactification limit $R\to \infty$, the expected
pure area exponentiation. This behaviour can be understood
as due to the sum over all instanton sectors. If only the
zero instanton sector is considered, in the decompactification limit
one exactly recovers the sum of the perturbative series in which
the light-cone gauge Yang-Mills propagator is prescribed according to 
Wu-Mandelstam-Leibbrandt. When instantons are disregarded, 
no pure area exponentiation
occurs, the string tension is different and, in the large-$N$ limit,
confinement is lost.
\end{abstract}
\vskip 2.0truecm
Padova preprint DFPD 98/TH 24.


\vskip 3.0truecm
\vfill\eject

\narrowtext
\section{Introduction}
\noindent
Quantum Yang-Mills theories on compact two-dimensional surfaces have been
extensively studied in the past years: in spite of their seeming
triviality, many interesting (and non-trivial) results were obtained
exploiting their non-perturbative solvability. In particular a string
picture, in the large-$N$ limit, was derived in \cite{Wati}, while
partition function \cite{Blau}, Wilson loops \cite{Bulo} and
field-strenght correlators \cite{Schniz} were computed exactly on arbitrary
genus. 

A further intriguing aspect is the appearance, on genus zero and
in the limit of large $N$, of a third order phase transition at a critical
value of $g^2NA$ \cite{Kaza}
($g^2$ being the Yang-Mills coupling costant and $A$ the
area of the sphere): a strong coupling phase, where a pure area exponentiation
for Wilson loops  dominates in the large-$A$ limit, 
is distinguished from a weak coupling phase with no
confining behaviour \cite{Lupi}. A clear physical picture of this phenomenon 
was presented in
\cite{Grosso}, showing, by explicit computations, that instanton
contributions are suppressed in the second case, while playing a
preminent role in driving the theory in the strong (confining) phase. 
The relevance of topologically non-trivial configurations
in connection with the Douglas-Kazakov phase transition was
also noticed some time before in ref.\cite{Case1}.

At the first sight this could appear as a paradox: confinement in $QCD_2$
is often regarded as a perturbative feature. As a matter of fact
the area
exponentation is simply obtained (for any $N$) by summing on the plane the
perturbative series, with the t'Hooft-CPV prescription for 
the gluon exchange potential \cite{Hooft}. 
Alternatively, from the results of \cite{Blau,Bulo}, 
one can easily realize that the very same result is
obtained for any value of $N$ in the decompactification limit when 
all instanton sectors are taken into account.

On the other hand an exact resummation of the perturbative series has
also been recently done \cite{Stau} adopting instead the
Wu-Mandelstam-Leibbrandt (WML) \cite{WML}
prescription for the gluon propagator, 
generalizing to all orders the ${\cal
O}(g^4)$
computation of \cite{noi1,noi2}; it leads, as firstly noticed in
\cite{noi1},
to a result different from a pure area-law exponentation
(which would be expected from the area-preserving diffeo-invariance of the
theory plus positivity arguments), and, in particular, 
predicting a different value for
the string tension.

Dramatically, the large-$N$ limit exhibits a non-confining behaviour, while
one easily realizes that, on the plane, the theory should be in the strong
coupling phase (the plane being thought as decompactification of a large
sphere).

\smallskip

In this
letter we discuss the reasons of these discrepancies, and show how the
WML computations presented in \cite{Stau,noi1,noi2} are indeed
perturbatively
correct, in the sense that what is missing exactly represents the instanton
contribution. This contribution can be eventually recovered by expanding 
the
functional integral as a sum over a class of topologically charged field
configurations. In so doing the usual expected area-law behaviour is
reproduced.

\section{The instanton expansion}
\noindent
Our starting point are the well-known expressions \cite{Blau} of the
exact partition function and of a non self-intersecting Wilson loop
for a pure $U(N)$ Yang-Mills theory on a sphere $S^2$ with area $A$
\begin{equation}
\label{partition}
{\cal Z}(A)=\sum_{R} (d_{R})^2 \exp\left[-{{g^2 A}\over 2}C_2(R)\right],
\end{equation}
\begin{equation}
\label{wilson}
{\cal W}(A_1,A_2)={1\over {\cal Z}N}\sum_{R,S} d_{R}d_{S}
\exp\left[-{{g^2 A_1}\over 2}C_2(R)-{{g^2 A_2}\over 2}C_2(S)\right]
\int dU {\rm Tr}[U]\chi_{R}(U) \chi_{S}^{\dagger}(U),
\end{equation}
$d_{R\,(S)}$ being the dimension of the irreducible representation $R(S)$ of
$U(N)$; $C_2(R)$ ($C_2(S)$) is the quadratic Casimir, $A_1+A_2=A$ are the
areas singled out by the loop, the integral in (\ref{wilson}) is over the
$U(N)$ group manifold while $\chi_{R(S)}$ is the character of the group
element $U$ in the $R\,(S)$ representation. 

Eqs. (\ref{partition}),
(\ref{wilson}) can be easily deduced from the solution of Yang-Mills theory 
on the cylinder, using the fact that the hamiltonian
evolution is governed by the laplacian on $U(N)$: we call eqs.
(\ref{partition}),(\ref{wilson}) the
heat-kernel representations of ${\cal Z}(A)$ and ${\cal W}(A_1,A_2)$,
respectively. 

On the other hands, as first noted by Witten \cite{Witte}, it is possible to
represent ${\cal Z}(A)$ (and consequently ${\cal W}(A_1,A_2)$) as a sum over
instable instantons, where each instanton contribution is 
associated to a finite,
but not trivial, perturbative expansion. The easiest way to see it, is 
to perform a Poisson resummation in eqs. (\ref{partition}),(\ref{wilson}).
 
To this purpose we write them explicitly for $N>1$ in the form
\begin{equation}
\label{partip}
{\cal Z}(A)=\frac{1}{N!}\exp \left [ -\frac{g^2 A}{24}N(N^2-1)\right ]
\sum_{m_i = -\infty}^{+\infty}\Delta^2(m_1,...,m_N)\exp\left [ 
-\frac{g^2A}{2}\sum_{i=1}^N (m_i-\frac{N-1}{2})^2\right],
\end{equation}
\begin{eqnarray}
\label{wilsonp}
&&{\cal W}(A_1,A_2)=\frac{1}{{\cal Z}N}\exp \left[-\frac{g^2A}{24}N(N^2-1)
\right ]\frac{1}{N!} \sum_{k= 1}^{N}\sum_{m_i=-\infty}^{+\infty}
\Delta(m_1,...,m_N)\times\\
&&\Delta(m_1+\delta_{1,k},...,m_N+\delta_{N,k})\ 
\exp\left [-\frac{g^2 A_1}{2}\sum_{i=1}^N (m_i-\frac{N-1}{2})^2
 -\frac{g^2 A_2}{2}\sum_{i=1}^N (m_i-\frac{N-1}{2}+\delta_{i,k})^2\right].
\nonumber
\end{eqnarray}
We have described the generic irreducible representation by means
of the set of integers $m_{i}=(m_1,...,m_{N})$, related to the
Young tableaux, in terms of which
we get
\begin{eqnarray}
\label{casimiri}
C_2(R)&=&\frac{N}{12}(N^2-1)+\sum_{i=1}^{N}(m_{i}-\frac{N-1}{2})^2,\nonumber
\\
d_{R}&=&\Delta(m_1,...,m_{N}).
\end{eqnarray}
$\Delta$ is the Van der Monde determinant and
 the integration in eq.(\ref{wilson})
has been performed explicitly, using the well-known formula for the 
characters in terms of the set $m_{i}$.

The instanton representation of ${\cal Z}(A)$ and of ${\cal W}(A_1,A_2)$ 
is now simply obtained \cite{Case2,Poli} by performing a Poisson resummation 
over $m_{i}$ 
\begin{eqnarray}
\label{poisson}
&&\sum_{m_{i}=-\infty}^{+\infty}F(m_1,...,m_{N})=
\sum_{n_{i}=-\infty}^{+\infty}\tilde{F}(n_1,...,n_{N}),\nonumber\\
&&\tilde{F}(n_1,...,n_{N})=\int_{-\infty}^{+\infty}dz_1...dz_{N}
\exp \left[2\pi i(z_1 n_1+...+z_{N}n_{N})\right]F(z_1,...,z_{N})
\end{eqnarray}
for the eqs.(\ref{partip},\ref{wilsonp}).

We have carefully repeated the original computations of ref.\cite{Grosso},
paying particular attention to the numerical factors and to the area
dependences; as a matter of fact, at variance with \cite{Grosso}, 
where interest
was focussed on the large-$N$ limit, we are mainly concerned with 
decompactification (large $A$) and with a comparison with the results
of ref.\cite{Stau} for any value of $N$. We have obtained
\begin{eqnarray}
\label{istanti}
{\cal Z}(A)&=&C(g^2 A,N)\sum_{n_{i}=-\infty}^{+\infty}
\exp\left[-S_{inst}(n_{i})\right]Z(n_1,...,n_{N}),\nonumber\\
{\cal W}(A_1,A_2)&=&\frac{1}{{\cal Z}N}C(g^2 A,N)\exp \left[
-g^2\frac{A_1A_2}{2A}\right]\sum_{n_{i}=-\infty}^{+\infty}
\exp\left[-S_{inst}(n_{i})\right]\nonumber\\
&\times&\sum_{k=1}^{N}\exp\left[-2 \pi i n_{k}\frac{A_2}{A}\right]
W_{k}(n_1,...,n_{N}),
\end{eqnarray}
where
\begin{eqnarray}
\label{quantita`}
C(g^2 A,N)&=& (i)^{N(N-1)}{{(g^2 A)^{-N^2/2}}\over {N!}}\exp
\left[-\frac{g^2 A}{24}N(N^2-1)\right]\nonumber\\
S_{inst}(n_{i})&=&\frac{2\pi^2}{g^2 A}\sum_{i=1}^{N}n_{i}^2,
\end{eqnarray}
and
\begin{eqnarray}
\label{zetawu}
Z(n_1,...,n_{N})&=&\exp(i\pi (N-1)\sum_{i=1}^{N}n_{i})\int
_{-\infty}^{+\infty}dz_1...dz_{N}\exp\left[-\frac{1}{2}
\sum_{i=1}^{N}z_{i}^2\right]\nonumber\\
&\times&\prod_{i<j}^{N}\Big(\frac
{4 \pi^2}{g^2 A}(n_{i}-n_{j})^2-(z_{i}-z_{j})^2\Big),\nonumber\\
W_{k}(n_1,...,n_{N})&=&\exp(i\pi (N-1)\sum_{i=1}^{N}n_{i})\int
_{-\infty}^{+\infty}dz_1...dz_{N}\exp\left[-\frac{1}{2}
\sum_{i=1}^{N}z_{i}^2\right]\times\nonumber\\
\prod_{i<j}^{N}\Bigl[\Big(\frac
{2 \pi}{\sqrt{g^2 A}}(n_{i}-n_{j})&+&ig^2\frac{A_2-A_1}{2\sqrt{g^2 A}}
(\delta_{i,k}-\delta_{j,k})\Big)^2
-\Big((z_{i}-z_{j})+i\frac{\sqrt{g^2 A}}{2}(\delta_{i,k}-\delta_{j,k})\Big)
^2\Bigr].
\end{eqnarray}

These formulae have a nice interpretation in terms of instantons.
Indeed, on $S^2$, there are non trivial solutions of the Yang-Mills equation,
labelled by the set of integers $n_{i}=(n_1,...,n_{N})$ 
\begin{equation}
\label{monopolo}
A_{\mu}(x)=\left(\matrix{n_1 {\cal A}_{\mu}^{0}(x) & 0 & \ldots & 0 \cr
                         0      & n_2 {\cal A}_{\mu}^{0}(x) & \ldots &0\cr
                         0 &\ldots&\ldots&0\cr
                         \vdots&\vdots&\ddots&\vdots\cr
                         0& 0 &\ldots &n_N{\cal A}_{\mu}^{0}(x)\cr
}\right)
\end{equation}
where ${\cal A}_{\mu}^{0}(x)={\cal A}_{\mu}^{0}(\theta, \phi)$ is the Dirac
monopole potential,
$${\cal A}_{\theta}^{0}(\theta, \phi)=0 , \quad\   {\cal A}_{\phi}^{0}
(\theta, \phi)={1-\cos \theta\over 2}, $$
$\theta$ and $\phi $ being the polar (spherical) coordinates on $S^{2}$.

The integer nature of the coefficients is of course a consequence of Dirac 
quantization condition or (more mathematically) of the fact that the original
$U(N)$-bundle has been reduced to a non-trivial $N$-torus bundle 
(see \cite{Witte,Blau2} for details). In the light of the above considerations,
eqs.(\ref{istanti}) can be interpreted as follows: 
$S_{inst}(n_i)$ represents the
classical action evaluated on the non-trivial solutions (\ref{monopolo}),
the exponential factor inside the sums is their contributions to
the Wilson loop, while $Z(n_{i})$ and $W_{k}(n_{i})$ are the quantum 
corrections, as anticipated by Witten using the Duistermaat-Heckman theorem; a
direct path-integral evaluation is presented in \cite{Blau2}.

>From the above representations it is rather clear why the decompactification
limit $A\to \infty$ should not be performed too early. Indeed on the plane
it is not easy to distinguish fluctuations around the instanton solutions
from Gaussian fluctuations around the trivial field configuration, since
$S_{inst}(n_{i})$ goes to zero for any finite set $n_{i}$ when $A\to
\infty$. For finite $A$ and finite $n_{i}$ instead, 
in the limit $g\to 0$, only the zero
instanton sector can survive in the Wilson loop expression (notice that
the power-like singularity $(g^2)^{-N^2/2}$ in the coefficient $C(g^2 A,N)$
exactly cancels in the normalization). In this limit each instanton 
contribution is ${\cal O}(\exp(-\frac{1}{g^2}))$; therefore instantons 
become crucial only when they are completely resummed.

On the other hand the zero instanton contribution should be obtainable 
in principle 
by means of perturbative calculations.  

In the following we compute from eqs.(\ref{istanti}) the exact
expression on the sphere $S^2$  of the zero instanton contribution
to the Wilson loop, obviously normalized to zero instanton 
partition function.

\section{Relation with perturbation theory}
\noindent
We write eq.(\ref{istanti}) for the zero instanton sector $n_{i}=0$.
Thanks to its symmetry, we can always choose $k=1$ and the equation becomes
\begin{eqnarray}
\label{zeroinst}
{\cal W}_0 &=&(2\pi)^{-\frac{N}{2}}\prod_{n=0}^{N}\frac{1}{n!}
\exp\left[-g^2\frac{A_1A_2}{2A}\right]
\int_{-\infty}^{+\infty}dz_1...dz_{N}\exp\left[-\frac{1}{2}\sum_{i=1}^{N}
z_{i}^2\right]\nonumber\\
&\times&\prod_{j=2}^{N}\Big[(z_1-z_{j})^2+i\sqrt{g^2A}(z_1-z_{j})-
g^2\frac{A_1A_2}{A}\Big]\Delta^2(z_2,...,z_{N}).
\end{eqnarray}
We introduce the two roots of the quadratic expression in the integrand
$z_{\pm}=z_1+i\alpha\pm i\beta$ with $\alpha=\frac{\sqrt{g^2 A}}{2}$
and $\beta=\frac{\sqrt{g^2}(A_1-A_2)}{2\sqrt{A}}$.
The previous equation then becomes
\begin{eqnarray}
\label{zetapiu}
{\cal W}_0 &=&(2\pi)^{-\frac{N}{2}}\prod_{n=0}^{N}\frac{1}{n!}
\exp\left[-g^2\frac{A_1A_2}{2A}\right]
\int_{-\infty}^{+\infty}dz_1...dz_{N}\exp\left[-\frac{1}{2}\sum_{i=1}^{N}
z_{i}^2\right]\nonumber\\
&\times&\Delta(z_{+},z_2,...,z_{N})\Delta(z_{-},z_2,...,z_{N}).
\end{eqnarray}
The two Van der Monde determinants can be expressed in terms of
Hermite polynomials \cite{Grosso} and then expanded in the usual way.
The integrations over $z_2,...,z_{N}$ can be performed, taking
the orthogonality condition into account; we get 
\begin{eqnarray}
\label{integrata}
{\cal W}_0 &=&(2\pi)^{-\frac{1}{2}}\prod_{n=0}^{N}\frac{1}{n!}
\exp\left[-g^2\frac{A_1A_2}{2A}\right]\prod_{k=2}^{N}(j_{k}-1)! 
\varepsilon^{j_1...j_{N}}\varepsilon_{j_1...j_{N}}\nonumber\\
&\times&\int_{-\infty}^{+\infty}dz_1\exp\Big[-\frac{z_1^2}{2}
\Big]He_{j_1-1}(z_{+})He_{j_1-1}(z_{-}).
\end{eqnarray}
Thanks to the relation
\begin{equation}
\label{laguerre}
\int_{-\infty}^{+\infty}dz_1\exp\Big[-\frac{z_1^2}{2}\Big]
He_{j_1-1}(z_{+})He_{j_1-1}(z_{-})=\sqrt{2\pi}(j_1-1)! L_{j_1-1}
(\alpha^2-\beta^2),
\end{equation}
we finally obtain our main result
\begin{equation}
\label{risultato}
{\cal W}_0=\frac{1}{N}\exp\left[-g^2\frac{A_1A_2}{2A}\right]\,
L_{N-1}^1(g^2\frac{A_1A_2}
{A}).
\end{equation}

At this point we remark that, in the decompactification limit $A\to
\infty, A_1$ fixed, the quantity in the equation above {\it exactly}
coincides, for any value of $N$, with eq.(11) of ref.\cite{Stau}, which
was derived following completely different considerations.
We recall indeed that their result was obtained by a full resummation
at all orders of the perturbative expansion of the Wilson loop
in terms of Yang-Mills propagators in light-cone gauge,
endowed with the WML prescription.

Several considerations can now be drawn.

First of all we notice that ${\cal W}_0$ does not exhibit the
usual area-law exponentiation;
actually, in the large-$N$ limit, exponentiation (and thereby confinement)
is completely lost, as first noticed in \cite{Stau}.
As a matter of fact, from eq.(\ref{risultato}), taking the limit 
$N\to \infty$, we easily get
\begin{equation}
\label{bessel}
\lim_{N\to \infty} {\cal W}_0=\sqrt{\frac{A_1+A_2}{\hat{g}^2 A_1A_2}}
J_1\Bigl(\sqrt{\frac{4\hat{g}^2A_1A_2}{A_1+A_2}}\Bigr)
\end{equation}
with $\hat{g}^2=g^2 N$.
At this stage, however, this is no longer surprising since 
${\cal W}_0$ does not contain any genuine non perturbative
contribution, {\it viz} instantons. If on the sphere $S^2$ we
consider the weak coupling phase $g^2 N A <\pi^2$, instanton
contributions are suppressed. As a matter of fact, eq.(\ref{bessel})
provides the complete Wilson loop expression in the  weak coupling
phase \cite{Kaza,Lupi}. In turn confinement occurs in
the strong coupling phase \cite{Grosso}.

For any value of $N$ the pure area law exponentiation follows, after
decompactification, from
resummation of all instanton sectors, changing completely the zero
sector behaviour and, in particular, the value of the string tension.

In the light of the considerations above, there is no contradiction between
the use of the WML prescription in the light-cone propagator and the pure
area law exponentiation; this prescription is correct but the ensuing
perturbative calculation can only provide us with the expression 
for ${\cal W}_0$. The paradox of ref.\cite{Stau} is solved
by recognizing that they did not take into account the genuine ${\cal
O}(\exp(-\frac{1}{g^2}))$ non
perturbative quantities coming, after decompactification, from the
instantons on the sphere.

What might instead be surprising in this context is the fact that, using the 
istantaneous 't Hooft-CPV potential and just
resumming at all orders the
related perturbative series, one still ends up with the
correct pure area exponentiation. It can perhaps be  naively understood 
if the exchange is interpreted
as a simple ``instantaneous'' increasing potential between a $q\bar q$ pair,
giving rise to hadronic strings in a natural way.
This  feature is likely
to be linked to some peculiar properties of the light-front vacuum
(we remind the reader that the light-cone CPV prescription follows
from canonical light-front quantization \cite{noi3}; still
we know it is perturbatively
unacceptable in higher dimensions \cite{Cappe} and cannot be smoothly
continued to any Euclidean formulation).

As a final remark we notice that, for $N=1$, we always find
the pure area exponentiation in the decompactification limit. 
This can be  understood by realizing that $W_{k}(n)=1$
in this case and therefrom all instanton sectors 
provide equal contributions in this limit. Still 't Hooft and WML
prescriptions lead to the same final result in a non trivial way
as only planar diagrams contribute in the first case.

\vskip 1.0truecm
${\bf Acknowledgement}$

One of us (L. G.) thanks D. Seminara for inspiring discussions
and the Center of Theoretical Physics at M.I.T. for hospitality.

\vfill\eject

\begin{references}
\bibitem{Wati}{ D.J. Gross and W. Taylor, Nucl. Phys. {\bf B400}, 181
(1993); {\it ibid.}{\bf B403}, 395 (1993).}
\bibitem{Blau}{ A.A. Migdal, Sov. Phys. JETP {\bf 42}, 413 (1975);\\
B.E. Rusakov, Mod. Phys. Lett. {\bf A5}, 693 (1990).}
\bibitem{Bulo}{ M.Blau and G. Thompson, Int. J. Mod. Phys. {\bf A7}, 3781 
(1992).}
\bibitem{Schniz}{ J. P. Nunes and H.J. Schnitzer, Int. J. Mod. Phys.
{\bf A12}, 4743 (1997).}
\bibitem{Kaza}{ M.R. Douglas and V.A. Kazakov, Phys. Lett. {\bf B319},
219 (1993).}
\bibitem{Lupi}{ D.V. Boulatov, Mod. Phys. Lett. {\bf A9}, 365 (1994);\\
J-M. Daul and V.A. Kazakov, Phys. Lett. {\bf B335}, 371 (1994).}
\bibitem{Grosso}{ D. J. Gross and A. Matytsin, Nucl. Phys. {\bf B429},
50 (1994); {\it ibid.} {\bf B437}, 541 (1997).}
\bibitem{Case1}{M. Caselle, A. D'Adda, L. Magnea and S. Panzeri,
{\it Two dimensional QCD on the sphere and on the cylinder}, in 
Proceedings of Workshop on High Energy Physics and 
Cosmology, (Trieste 1993) eds. E.Gava, A.Masiero, K.S.Narain, S.Randjbar-Daemi 
and Q.Shafi, World Scientific, Singapore, 1994.}
\bibitem{Hooft}{ G. 't Hooft, Nucl. Phys. {\bf B75}, 461 (1974).}
\bibitem{Stau}{ M. Staudacher and W. Krauth, Phys. Rev. {\bf D57},
2456 (1998).}
\bibitem{WML}{ T.T. Wu, Phys. Lett. {\bf 71B}, 142 (1977);\\ S.
Mandelstam, Nucl. Phys. {\bf B213}, 149 (1983);\\
G. Leibbrandt, Phys. Rev. {\bf D29}, 1699 (1984).}
\bibitem{noi1}{ A. Bassetto, F. De Biasio and G. Griguolo, Phys. Rev.
Lett. {\bf 72}, 3141 (1994).}
\bibitem{noi2}{ A. Bassetto, D. Colferai and G. Nardelli, Nucl. Phys.
{\bf B501} 227 (1997) and Erratum-{\it ibid.} {\bf B507}, 746 (1997).}
\bibitem{Witte}{ E. Witten, Commun. Math. Phys. {\bf 141}, 153 (1991) and
J. Geom. Phys. {\bf 9}, 303 (1992).}
\bibitem{Case2}{M. Caselle, A. D'Adda, L. Magnea and S. Panzeri,
Nucl. Phys. {\bf B416}, 751 (1994).}
\bibitem{Poli}{ J.A. Minahan and A. P. Polychronakos, Nucl. Phys.
{\bf B422}, 172 (1994).}
\bibitem{Blau2}{ M. Blau and G. Thompson, J. Math. Phys. {\bf 36}, 2192
(1995).}
\bibitem{noi3}{ A. Bassetto, G. Nardelli and R. Soldati, {\it
Yang-Mills Theories in Algebraic Non-Covariant Gauges}, World
Scientific, Singapore 1991.}
\bibitem{Cappe}{ D.M. Capper, J.J. Dulwich and J. Litvak, Nucl. Phys.
{\bf B241}, 463 (1984).}
\end{references}
\end{document}